
\input phyzzx
 \hsize=6.5in
\hoffset=0.0in
\voffset=0.0in
\vsize=8.9in
\FRONTPAGE
\line{\hfill BROWN-HET-891}
\line{\hfill February 1993}
\bigskip
\centerline{\bf A COSMOLOGICAL THEORY WITHOUT SINGULARITIES}
\bigskip
\centerline{R. Brandenberger$^{1)}$ , V. Mukhanov$^{2 , \ast)}$ and
 A. Sornborger$^{1)}$}
\medskip
\centerline{$^{1)}$  Physics Department }
\centerline{Brown University }
\centerline{Providence, RI  02912, USA }
\medskip
\centerline{$^{2)}$  Institut f\"ur Theoretische Physik }
\centerline{ETH Z\"urich, H\"onggerberg }
\centerline{CH--8093 Z\"urich, Switzerland }
\bigskip
\centerline{Abstract}
A theory of gravitation is constructed in which all homogeneous and isotropic
solutions are nonsingular, and in which all curvature invariants are bounded.
All solutions for which curvature invariants approach their limiting values
approach de Sitter space.  The action for this theory is obtained by a higher
derivative modification of Einstein's theory.  We expect that our model can
easily be generalized to solve the singularity problem also for anisotropic
cosmologies.
\par
\vfill
\noindent
 $\ast$ On leave of absence from Institute for Nuclear Research, Academy of
Sciences, 117 312 Moscow, Russia.
\endpage
\chapter{Introduction}
\par
One of the outstanding problems in the theory of gravitation (and more
generally in the quest for a unified theory of all interactions) is the
singularity problem.  According to the Penrose--Hawking theorems$^{1)}$,
general relativity (GR) manifolds are, in general, geodesically incomplete,
which is a sign that singularities in space--time occur.
\par
Singularities are undesirable for a theory which claims to be complete, since
their existence implies that space--time cannot be continued past them.  The
space--time structure becomes unpredictable already at the classical level.
\par
Two important examples of singularities in GR are the initial and final
singularities in a closed Universe and the singularity in the center of the
black hole.  In the former case, the singularity implies we cannot answer the
question what will happen after the ``big crunch'', or (in the case of an
expanding Universe) what was before the ``big bang''.
\par
The presence of singularities is an indication that GR is an incomplete theory.
 Wheeler even talks about a ``crisis in physics''$^{2)}$.  It is a widespread
opinion that either quantum gravity or a more fundamental theory such as string
theory will provide a cure for the ``sickness" of GR.  However, quantum gravity
does not yet exist as a self--consistent non--perturbative theory.  Neither
does string theory exist as a unique theory capable of addressing the
singularity problem of gravity in a definitive way, although interesting
string--specific ideas have recently been put forwards$^{3)}$.
\par
Because of the absence of a completely developed fundamental theory on the
basis of which we could address the singularity problem, we will use a rather
different approach.  Any fundamental theory will, in the region of low
curvature, give an effective action for a four--dimensional space--time metric
$g_{\mu v}$ which to lowest order must agree with the Einstein action.  We will
try to construct (guess) an effective action for $g_{\mu v}$ which solves the
singularity problem and which in the low curvature limit reduces to the
Einstein action.  It is possible that in such a manner we will be able to
discover important features of the future fundamental theory.  We might also
gain information which will help in finding this fundamental theory.
\par
Before discussing the ideas behind our construction of the effective action for
gravity, we return to the Penrose--Hawking theorems$^{1)}$.  They do not give
us any detailed information about the nature of the singularity.  However, in
the two examples discussed above -- collapsing Universe and black hole -- we
know that at the singularity some of the physically measurable curvature
invariants like $R$, $R_{\mu \nu}\>R^{\mu \nu}\, ,\>  C^2 = C_{\alpha \beta
\gamma \delta}\>C^{\alpha \beta \gamma \delta}$ diverge (here, $R$ is the Ricci
scalar, $R_{\mu \nu}$ the Ricci tensor, and $C_{\alpha \beta \gamma \delta}$
the Weyl tensor).  It is reasonable to assume that the divergence of some
curvature  invariants at the singularity  is a fairly general phenomenon..  In
fact, for singularities reached on timelike curves in a globally hyperbolic
space--time it can be proved$^{4)}$ that the Riemann tensor becomes infinite.
Hence, as a first step we will find a mechanism to bound all the curvature
invariants.
\par
Limitation principles play a very important role in physics.  Special
relativity includes as one of its fundamentals the principle that no particle
velocity can exceed the speed of light.  The cornerstone of quantum mechanics
is the uncertainty principle which states that the second fundamental constant,
Planck's constant $\hbar$, gives the minimal phase space volume a particle can
be localized in.  The third fundamental constant, Newton's gravitational
constant $G$, has not yet been used in any limitation principle.
\par
Thus it is natural to assume that there exists a fundamental length
$\ell_{p\ell} \sim \left ( G\hbar c^{-3}\right )^{1/2} \simeq 10^{-33}$ cm in
nature (determined by $G$) such that there is no curvature corresponding to
scales $\ell < \ell_{p\ell}$.  There are strong indications that this will in
fact arise in quantum gravity$^{5)}$ or string theory$^{3)}$.  From the
existence of a fundamental length it follows by simple dimensional
considerations that all curvature invariants are limited:
$$
\vert R\vert \leq \ell_{p\ell}^{-2}\>,\>\vert R_{\mu \nu} R^{\mu \nu} \vert
\leq \ell_{p\ell}^{-4}\>,\>\vert C_{\alpha \beta \gamma \delta}\> C^{\alpha
\beta \gamma \delta} \vert \leq \ell_{p\ell}^{-8}\> ,\dots\eqno(1.1)
$$
To realize the idea of a fundamental length it is necessary to construct a
theory in which all curvature invariants are bounded.  Since there are an
infinite number of curvature invariants and since bounds on low order
invariants do not necessarily imply bounds on higher order invariants, it is a
rather formidable task to construct such a theory.
\par
Fortunately we can simplify the problem drastically by making use of the
``limiting curvature hypothesis'' (LCH) construction$^{6)}$, according to which
one looks for a theory in which:
\item{i)}A finite number of invariants  are bounded by an explicit construction
(e.g. $\vert R\vert \leq \ell_{p\ell}^{-2}$ and $\vert R_{\mu \nu} R^{\mu
\nu}\vert \leq \ell_{p \ell}^{-4}$).
\item{ii)} When these invariants take on their limiting values, any solution of
the field equations reduces to a definite nonsingular solution (e.g. de Sitter
space).

\noindent In this case it follows automatically that all curvature invariants
are limited.
\par
The LCH incorporates  in a natural manner Penrose's hypothesis$^{7)}$ that the
Weyl tensor $C$ should vanish at the beginning (and end) of the Universe.  This
follows since by the LCH the Universe near the big bang and big crunch is de
Sitter and that $C=0$ for a de Sitter Universe.
\par
A theory in which the LCH is realized has some attractive features, both for
cosmology$^{6)}$ and for black holes$^{8)}$.   In cosmology, the present
homogeneous expanding Universe would have started out with a de Sitter phase.
In this case, we would have some (maybe unusual) realization of the oscillating
Universe scenario.  Entropy considerations tell us that only for a perfectly
homogeneous and isotropic Universe could we have perfect periodicity.  In
general, we must have a nontrivial realization.  Including inhomogeneities, we
might obtain a multi--Universe model in which one collapsing Universe splits
into several de Sitter bounces.
\par
For black holes, the LCH gives the attractive picture that inside of the
horizon instead of a singularity at the center we would have a piece of a de
Sitter Universe  which could be the source of other Friedmann (baby) Universes
(see Fig. 1).  In this case, the difficult question$^{9)}$ concerning
information loss when matter falls into a black hole have a natural answer:
the information which is lost to the observer external to the Schwarzschild
horizon is stored in the baby Universe.  In addition, using this picture
provides a good starting point to attack the issue of the final stage of an
evaporating black hole, a problem which has recently been of high interest in
the context of two--dimensional quantum gravity$^{10)}$.
\par
In this paper we construct an effective action for gravity in which all
homogeneous and isotropic solutions are nonsingular and at high curvature
approach de Sitter space.  (A brief summary of our work was published in Ref.
11.) In order to implement the LCH we procede in analogy to a technique by
which  point particle velocities can be limited, thus  achieving the transition
between Newtonian mechanics and point particle motion in special relativity
(SR) (see also Ref. 12).  An extension of our construction to inhomogeneous
cosmologies and to black hole metrics will be presented separately.$^{13)}$
\par
In the following section we present the general theory of how to implement the
LCH.  We obtain a fairly general effective action for gravity as a higher
derivative modification of the Einstein action, specialize to the case of an
isotropic, homogeneous Universe and derive the resulting equations of motion.
\par
In Section 3 we analyze a simple model which yields a nonsingular Universe
without limiting curvature.  We discuss the effects of including spatial
curvature (i.e. $k \neq 0$) and hydrodynamical matter.  The analysis of the
more complicated  model with limiting curvature is given in Section 4.  Section
5 contains conclusions and further discussion.
\chapter{Theory}
\par
In order to realize the LCH and hence to avoid singularities it is necessary to
abandon at least one of the key assumptions on which the Penrose--Hawking
theorems are based.  The two most important assumptions are:
\item{i)} The energy dominance condition, a simplified version of which
appropriate for cosmology is $\epsilon > 0$ and $\epsilon + 3 p \geq 0$, where
$\epsilon$ and $p$ are matter energy density and pressure respectively.
\item{ii)} The Einstein equations are universally true.
\par
There is no reason to believe that these assumptions will be valid at very high
energies and curvatures.  First of all, already in matter theories routinely
studied by particle physicists, the energy dominance condition is not always
true.  For example, the effective equation of state for a homogeneous, slowly
varying scalar field configuration with potential energy is $p\simeq -
\epsilon$, thus violating the energy dominance condition.  This matter
evolution scenario is in fact the basis for the inflationary Universe$^{14)}$.
\par
Note, however, that inflationary Universe models do not cure the problem of the
final singularity.  There may be nonsingular solutions for a collapsing
Universe filled with scalar field matter, but they are of measure zero.
Rather, in this case the typical solutions have an effective equation of state
$p = \epsilon$ (the kinetic term for the scalar field dominates), not $p =  -
\epsilon$, and hence have a final singularity.  Our goal is to construct a
theory in which \undertext{all} solutions are nonsingular.
\par
Concerning the second key assumption of the Penrose--Hawking theorems, it is
well known that the Einstein theory can only be the effective  theory of
gravity at low curvatures.  Perturbative quantum gravity calculations$^{15)}$,
vacuum polarization effects of quantum matter fields in an external
gravitational background$^{16)}$, and also considerations based on string
theory$^{17)}$ all show that the effective equations for the gravitational
field should be modified at higher curvatures.  In a perturbative analysis, the
modifications take the form of higher  derivative terms which are usually
important only at very high (Planck) curvatures.  Hence, provided the effective
action approach is valid at all at high curvatures, this effective action will
certainly not be of pure Einstein form.
\par
To summarize, there are two ways to modify the theory at high curvatures in
order to avoid singularities:
\item{i)} Modify the matter action by including terms which violate the energy
dominance condition.
\item{ii)} Modify the gravitational field equations.
\par
The first approach was explored in Ref. 18.  However, the weakness of this
approach is the absence of a good physical motivation for the modification.  In
addition, it seems impossible to avoid singularities associated with purely
gravitational modes which do not couple to matter.
\par
The second approach is much better motivated since higher derivative correction
terms to the Einstein action are predicted by many theories$^{15-17)}$.  Hence,
our starting point will be to look for an effective action for gravity of the
form
$$
S_g = -\, {1\over {16 \pi G}} \int F \left (R\, , R_{\mu \nu} R^{\mu \nu}\, ,
C_{\alpha \beta \gamma \delta} C^{\alpha \beta \gamma \delta}\, , \dots \right
) \> \sqrt{-g} \, d^4 x + {\rm nonlocal\> terms}\, ,\eqno(2.1)
$$
where the dots denote the dependence of $F$ on other curvature invariants.  At
low curvatures, the leading term in $F$ is simply $R$.
\par
The action (2.1) can be viewed as the effective action of some fundamental
theory such as quantum gravity or string theory.  In these theories we are at
present unable to calculate the nonperturbative effective action.  Hence, as
mentioned in the introduction, our approach  will be to construct (guess) an
effective action of the form (2.1) to obtain a theory in which \undertext{all}
solutions are nonsingular.
\par
To simplify  the considerations, we shall neglect nonlocal terms.  In our
approch, this is justifiable since if we are able to solve the singularity
problem in a purely local theory, we expect that the nonlocal terms (which are
inevitable for example because of particle production) will not drastically
change the asymptotic behavior of our theory because of its special properties
(see Section 5).
\par
Key to the analysis is the assumption about validity of the background field
approximation for the gravitational field up to high curvatures.  Such an
approximation will only be justified if the quantum fluctuations around this
metric are sufficiently small.  If the gravitational field is asymptotically
free at high curvatures (see Section 3), we can hope that this approach will be
valid.  As we shall see, there are features in our theory which indicate that
this will really be the case.
\par
For the moment we shall ignore matter (later we will show that the presence of
matter does not change the solutions at high curvatures in an important way).
Thus, our starting point is the effective action
$$
S_g = \,-\,{1\over {16 \pi G}} \int F\left (R\, , R_{\mu \nu}\, R^{\mu \nu}\, ,
C^2\, , \dots \right )\> \sqrt{- g}\, d^4 x\, .\eqno(2.2)
$$
\par
The usual Einstein theory in the absence of matter has only one solution --
Minkowski space -- for a homogeneous and isotropic Universe.  Any non--Einstein
theory of gravity gives rise to fourth (or higher) order equations of motion
and hence to a large number of cosmological solutions.  In general, the
singularity problems of such a theory are much worse than in Einstein gravity.
A simple example is $R^2$ gravity
$$
F(R) = R + \alpha R^2\eqno(2.3)
$$
which is conformally equivalent$^{19)}$ to Einstein gravity plus scalar field
matter and which hence has many isotropic singular solutions (even without
matter).  Thus, the theory we are looking for must be a very special higher
derivative gravity model.
\par
We wish to construct an effective action for gravity in which all homogeneous
and isotropic solutions are nonsingular and in which all curvature invariants
are limited (in Section 5 we will indicate how to extend our analysis to
anisotropic models$^{13)}$). To motivate our construction, it is useful to keep
in mind ways of writing the action for two well-known physical theories in
which certain physical quantities are bounded:  special relativity and the
Born--Infeld theory of electromagnetism$^{20)}$.
\par
To impose bounds on physical quantities in an explicit manner, it is convenient
to employ a Lagrange multiplier technique proposed by Altshuler$^{12)}$.  To
explain how this technique works, we first consider the simple example of point
particle motion.  We start with the action for a nonrelativistic particle of
mass $m$ and world line $x (t)$.  We demonstrate how to explicitly implement
the limitation on the particle velocity, and in particular how to obtain the
action for point particle motion in special relativity.  The nonrelativistic
action with which we start is
$$
S_{\rm old} = m \int dt\>{1\over 2}  {\dot x}^2\, .\eqno(2.4)
$$
In order to construct a new theory with bounded velocity, we introduce a
``Lagrange multiplier field'' $\phi (t)$, which couples to some function of the
quantity whose value we want to limit, and a potential $V(\phi)$ for this
field:
$$
S_{\rm new} = m \int dt \left [ {1\over 2} \>{\dot x}^2 + \phi\, {\dot x}^2 - V
(\phi)\right ]\eqno(2.5)
$$
Let us stress that $\phi$ is not a dynamical field.  Provided that $\partial
V/\partial \phi$ is bounded, the constraint equation (i.e. the variational
equation with respect to $\phi$) ensures that $\dot x$ is bounded.  In order to
obtain the correct Newtonian limit for small $\dot x$ and small $\phi\, ,\,
V(\phi)$ must be proportional to $\phi^2$ as $\vert \phi\vert \rightarrow 0$.
One of the simplest potentials which satisfies the above asymptotic conditions,
$$
V(\phi) = {2\phi^2\over {1+2\phi}}\, ,\eqno(2.6)
$$
leads to special relativity.  In fact, eliminating the Lagrange multiplier
using the constraint equation and substituting the result into (2.5) yields (up
to a constant term which does not effect the equations of motion) the
relativistic point particle action
$$
S_{\rm new} = m \int dt\>\sqrt{1-{\dot x}^2}\, .\eqno(2.7)
$$
\par
Let us return to the theory of gravitation.  In the notation of the above
example, the ``old'' theory will be given by the Einstein action.  In order to
implement the LCH, we wish to impose restrictions on some curvature invariants
$I_1\, , \, I_2\, ,\,\dots , I_n$ in an explicit manner.  The general form of a
higher derivative local modification of the Einstein action involving the
invariants $I_1\, ,\, \dots , I_n$ is
$$
S_g = -\>{1\over {16 \pi G}} \int \left [ R+F(I_1\, ,\, I_2\, ,\, \dots ,
I_n)\right ] \>\sqrt{-g}\, d^4 x\, ,\eqno(2.8)
$$
where $F$ is some function of the invariants $I_1\, ,\, \dots\, , I_n$.
\par
By introducing Lagrange multiplier fields $\phi_1 (t)\, ,\, \dots , \phi_n
(t)$, the above action can be rewritten as
$$
S_g = -\>{1\over {16\pi G}} \int \left [ R+\phi_1 f_1 (I_1) + \dots + \phi_n
f_n (I_n) + V(\phi_1 \dots \phi_n)\right ]\>\sqrt{-g} d^4 x\, ,\eqno(2.9)
$$
where $f_i (I_i)$ are functions we can choose as we want. The actions (2.8) and
(2.9) are equivalent provided that the potential $V(\phi_1\, ,\dots , \phi_n)$
satisfies the following partial differential equation
$$
-\, \sum_{i=1}^n \phi_i\> {\partial V\over {\partial \phi_i}} + V(\phi_1 \dots
\phi_n) = F\left(f_1^{-1} \left( {\partial V\over {\partial \phi_1}}\right )\,
, \dots f_n^{-1} \left({\partial V\over {\partial \phi_n}}\right )\right )\,
.\eqno(2.10)
$$
This follows immediately by using the constraint equations for (2.9)
$$
f_i (I_i) = {\partial V\over {\partial \phi_i}} \qquad i=1\, ,\dots , n\,
.\eqno(2.11)
$$
\par
We see from the constraint equations (2.11) that by appropriate choice of the
functions $f_i$ and $V$ we can implement bounds on the invariants $I_1\, ,
\dots , I_n$.   Variation of the action (2.9) with respect to $g_{\mu \nu}$
yields the field equations..
\par
First, we try to construct the simplest theory in which the LCH is realized.
At least for simple models (like the isotropic Universe) it is natural to
choose as one of the invariants
$$
I_1 = R - \sqrt{3}\> ({4R_{\mu \nu}\,  R^{\mu \nu} - R^2})^{1/2}\eqno(2.12)
$$
since for a homogeneous, spatially flat Universe it is equal to $12 H^2$.  This
invariant will be used to limit the curvature by some (e.g. Planckian) value.
The second invariant $I_2$  we will take on such a form as to implement in the
theory the condition that in the asymptotic regions all of the solutions evolve
to de Sitter.  The simplest way to do this is to pick $I_2$ such that $I_2 = 0$
only for de Sitter space (Minkowski space is included as a special case), and
to make sure that
$$
I_2 \rightarrow 0 \quad {\rm as} \quad \vert \phi_2 \vert \rightarrow \infty\,
.\eqno(2.13)
$$
\par
For homogeneous and isotropic space--times, it can be shown that
$$
I_2 = 4 R_{\mu \nu}\>R^{\mu \nu} - R^2\eqno(2.14)
$$
is a good choice, since $I_2 = 0$ only for de Sitter space. Note that in
general, $I_2$ is positive semidefinite.  However, for inhomogeneous and
anisotropic space--times (e.g. when $C^2 \neq 0$), the above form of $I_2$ is
insufficient to single out de Sitter space as an asymptotic solution.  This is
obvious from considering the Schwarzschild  metric for which $I_2 = 0$. Hence,
in the general case we$^{13)}$ should add to (2.14) terms which depend on $C^2$
and vanish for conformally flat space--times.
\par
However, for a homogeneous and isotropic Universe it is (as we will show)
sufficient to consider the action in the following general form:
$$
S_g = -\, {1\over {16 \pi G}} \int \left [ R+\phi_1\, f_1\, (I_1) +\phi_2\,
f_2\,(I_2) + V(\phi_1\, , \phi_2)\right ] \>\sqrt{-g}\>d^4 x\, .\eqno(2.15)
$$
The variational field equations which follow from (2.15) are
$$
\eqalign{
R_\beta^\alpha &- {1\over {2}} \delta_\beta^\alpha\, R - {1\over 2}
\delta_\beta^\alpha\, V = \left (\phi_1\>{\partial f_1\over {\partial
I_1}}\right )_{, \beta}^{, \alpha} \, - \left (\phi_1\> {\partial f_1\over
{\partial I_1}}\right )_{, \sigma}^{, \sigma}\cr
&- \phi_1\> {\partial f_1\over {\partial I_1}}\> R_\beta^\alpha + {1\over 2}
\phi_1\, f_1\, \delta_\beta^\alpha - 4\, \left (\phi_2\, {\partial f_2\over
{\partial I_2}}\, R^{\sigma\tau}\right )_{, \sigma\tau}\,
\delta_\beta^\alpha\cr
&- 4 \left (\phi_2 {\partial f_2\over {\partial I_2}}\, R_\beta^\alpha \right
)_{, \sigma}^{, \sigma} + 4 \left (\phi_2\, {\partial f_2\over {\partial
I_2}}\, R_\sigma^\alpha\right )_{, \beta}^{, \sigma} + 4 \left (\phi_2\,
{\partial f_2\over {\partial I_2}}\, R_\beta^\sigma\right )_{, \sigma}^{,
\alpha}\cr
&+2 \delta_\beta^\alpha \left (\phi_2\, {\partial f_2\over {\partial I_2}}
R\right )_{, \sigma}^{, \sigma}\> -2 \left (\phi_2\, {\partial f_2\over
{\partial I_2}} R\right )_{, \beta}^{, \alpha} - 8 \phi_2 \, {\partial f_2\over
{\partial I_2}}\, R^{\alpha \gamma}\, R_{\gamma \beta}\cr
&+ 2\phi_2 {\partial f_2\over {\partial I_2}} R\, R_\beta^\alpha +{1\over 2}
\delta_\beta^\alpha\, \phi_2\, f_2\, ,}\eqno(2.16)
$$
and the constraint equations are
$$
\eqalign{
f_1 (I_1) &= - {\partial V\over {\partial \phi_1}}\cr
f_2 (I_2) &= - {\partial V\over {\partial \phi_2}}\, .}\eqno(2.17)
$$
We will simplify the theory further by assuming a factorizable  potential
$$
V (\phi_1\, ,\, \phi_2) = V_1 (\phi_1) + V_2 (\phi_2)\, .\eqno(2.18)
$$
\par
The asymptotic conditions on the potentials $V_1$ and $V_2$ follow from
demanding that the theory reduces to the Einstein theory at small curvatures,
and that the LCH is realized.  The first condition yields
$$
V_i (\phi_i) \sim \phi_i^2 \quad \vert \phi_i \vert \ll 1\> ;\>i=1\, ,\, 2\,
.\eqno(2.19)
$$
In order to limit $R$ explicitly we can try a potential which to leading order
takes the form
$$
V_1 (\phi_1) \sim \phi_1 \qquad \vert \phi_1 \vert \gg 1\, ,\eqno(2.20)
$$
and to obtain de Sitter solutions in the asymptotic regions we need a potential
which at large $\phi_2$ increases less quickly than $\phi_2$.  We assume an
asymptotic form:
$$
V_2 (\phi_2) \sim\> {\rm const}\> \qquad \vert \phi_2 \vert \gg 1\,
.\eqno(2.21)
$$
In this case, provided $f_2 (I_2) \rightarrow 0$ as $I_2 \rightarrow 0$, the
constraint equation (2.17) implies that $I_2 \rightarrow 0$ as $\vert \phi_2
\vert \rightarrow \infty$, and we have a chance of realizing the LCH, provided
that the evolution of the scalar fields $\phi_1$ and $\phi_2$ is appropriate, a
question which needs detailed investigation.
\par
To conclude this section we will write down equations (2.16) and (2.17)
explicitly for a homogeneous and isotropic metric with scale factor $a(t)$ in
the contracting phase (i.e. $H < 0$):
$$
ds^2 = dt^2 - a^2 (t) \left ( {1\over {1-kr^2}} dr^2 + r^2 d\theta^2 + r^2
\sin^2 \theta\, d\phi^2\right )\, .\eqno(2.22)
$$
We choose simple functions $f_1$ and $f_2$:
$$
\eqalign{
f_1 (I_1) &= I_1\cr
f_2 (I_2) &= - \sqrt{I_2}}\eqno(2.23)
$$
Thus, our final action takes the form
$$
S_g = - {1\over {16 \pi G}} \int \left [ (1 + \phi_1) R - (\phi_2 + \sqrt{3}\,
\phi_1)\> \sqrt{4R_{\mu \nu} R^{\mu \nu} - R^2} + V_1 (\phi_1) + V_2
(\phi_2)\right ] \sqrt{-g} d^4 x\, .\eqno(2.24)
$$
\par
As is well known from the derivation of the Friedmann--Robertson--Walker
equations in Einstein gravity, the only independent equation of motion  is the
$0-0$ equation.  In our case, we have in addition the constraint equations
(2.17).  The full set of equations can be obtained by inserting the metric
(2.22) into (2.16) and (2.17).
\par
The resulting $\phi_1\, , \phi_2$ and $0-0$ equations are
$$
H^2 + {k\over {a^2}} = {1\over 12}\> V_1^\prime\eqno(2.25)
$$
$$
\dot H - {k\over {a^2}} = \, -\, {1\over {\sqrt{12}}}\> V_2^\prime\eqno(2.26)
$$
$$
-\, {1\over 2} (V_1 + V_2 ) + 3H^2 (1-2 \phi_1) + 3\, {k\over {a^2}} (4 \phi_1
+ 1) = \sqrt{3}\, H \left ({\dot \phi}_2 + 3 H \phi_2 - {k\over {Ha^2}}
\phi_2\right )\eqno(2.27)
$$
Another way to obtain the same equations is to substitute the ansatz (2.22)
with $g_{00} = N (t)^2$ into the action (2.24) and to vary with respect to $N\,
, \phi_1$ and $\phi_2$ (see e.g. Ref. 21).  Adding to the system matter with
action
$$
S_m = \int {\it L}_m \>\sqrt{-g}\, d^4 x\eqno(2.28)
$$
where ${\it L}_m$ is the matter Lagrangian, only leads to an additional term
$$
{{8 \pi}\over 3}\>  G\, \rho_m\eqno(2.29)
$$
on the left hand side of the $0-0$ equation.
\par
In the following sections, we shall show that all solutions of the above
equations are free of singularities.
\chapter{Nonsingular Universe without Limiting Curvature}
\par
Since our goal is primarily to construct a nonsingular Universe model and only
secondarily to limit the curvature,  we first consider a simple model in which
the $\phi_1$ field is absent.  In this case, it is easier to discuss our
techniques of analysis.
\par
We will show that for this model all solutions for a collapsing Universe are
nonsingular and asymptotically approach de Sitter solutions.  However, there is
no general (i.e. solution independent) bound on the effective cosmological
constant of the de Sitter period.
\par
In this section we set $\phi_2 \equiv \phi$ and $V_2 \equiv V$. The equations
of motion are given by (2.26) and (2.27).  Let us first consider a spatially
flat $(k=0)$ collapsing model without matter. In this case, the equations of
motion are
$$
{\dot H} = \, -\, {1\over {2\sqrt{3}}} V^\prime\eqno(3.1)
$$
$$
{\dot \phi} = \, -\, 3H\phi + \sqrt{3} H - {1\over {2\sqrt{3}\, H}} V\, .
\eqno(3.2)
$$
\par
The phase space of this model is the two dimensional $(\phi\, , H)$ plane.  The
phase space trajectories can be understood by considering $dH/d\phi$
(determined immediately from (3.1) and (3.2))
$$
{dH\over {d\phi}} = \, -\, {V^\prime\over {\sqrt{12}}} \> \left (- 3 H\phi +
\sqrt{3} H \, -\, {1\over {2\sqrt{3} H}}\> V \right )^{-1}\, .\eqno(3.3)
$$
\par
{}From (3.3), it follows that provided that $V(\phi)$ is bounded at large
$\phi$ (as postulated in (2.21)), then as $\phi$ tends to infinity, $H$
approaches a finite value, i.e. for any solution, the effective cosmological
constant in the large $\phi$ region is bounded.  In this case, it follows from
(3.2) that in a collapsing Universe, for large $\phi$
$$
\phi (t) \sim e^{3\vert H\vert t}\, .\eqno(3.4)
$$
Our choice of invariant  $I_2$ has led to the conclusion that the asymptotic de
Sitter solutions are attractor solutions.  This conclusion holds independent of
the specific choice of the potential $V(\phi)$, as long as the asymptotic
condition (2.21) is satisfied.
\par
{}From (3.4) it follows that all solutions for a contracting Universe are free
of singularities.  It takes infinite time to reach $\phi = \infty$.
\par
To concretize the consideration, we consider a simple potential which satisfies
the asymptotic conditions (2.19) and (2.21):
$$
V = \sqrt{12}\, H_0^2 {\phi^2\over {1+\phi^2}}\, ,\eqno(3.5)
$$
where $H_0$ is a constant (in the model with limiting curvature discussed in
Section 4, $H_0$ sets the scale of this limiting curvature).
\par
The phase space trajectories $(\phi (t)\, , H(t))$ in a collapsing Universe are
shown in Fig. 2.  The numerical results were obtained using the specific
potential  (3.5).  However, as discussed above, the main features of the
diagram depend only on the asymptotic properties.
\par
First we note that there is only one singular point $({\dot \phi} = {\dot H} =
0 )$ in the phase plane.  This point is
$$
(\phi\, ,\, H) = (0\, ,\, 0)\eqno(3.6)
$$
and corresponds to Minkowski space--time.
\par
There are two classes of trajectories which are
asymptotically de Sitter.  Those starting at large positive values of $\phi$ go
off to $\phi = \infty$, reaching their asymptotic value of $H$ from above (i.e.
${\dot H} < 0$). Those starting with large negative values of $\phi$ tend to
$\phi = - \infty$ with ${\dot H} > 0$.
\par
For small values of $H$ and $\phi$ we can use the asymptotic condition (2.19)
on $V(\phi)$ to conclude that there are periodic solutions about Minkowski
space.  In this limit, the basic equations (3.1) and (3.2) become
$$
{\dot H} = \, -\, {1\over {2 \sqrt{3}}}\> {\partial V\over {\partial \phi}}
\simeq -\, 2H_0^2 \phi\eqno(3.7)
$$
$$
{\dot \phi} \simeq {1\over {\sqrt{3}}}\> {3H^2 - {1\over 2} V\over { H}} \simeq
H_0 {\sqrt{3} (H / H_0)^2 - \phi^2\over {H / H_0}}\, ,\eqno(3.8)
$$
where for $V(\phi)$  we have inserted the general asymptotic form
$$
V (\phi) \simeq 2 \sqrt{3}\> H_0^2\> \phi^2 \, ,\eqno(3.9)
$$
valid for small $\phi$.  The numerical factor $2 \sqrt{3}$ has been inserted to
eliminate numerical constants in the following equations.
\par
It is convenient to introduce  a rescaled time
$$
\tau \equiv H_0 t\eqno(3.10)
$$
and a dimensionless measure of $H$:
$$
y \equiv H /H_0\eqno(3.11)
$$
With $d /d\tau$ denoted by a prime, equations (3.7) and (3.8) become
$$
\eqalign{y^\prime &= - 2 \phi\cr
\phi^\prime &= {\sqrt{3}\, y^2 - \phi^2\over {y}}}\eqno(3.12)
$$
To see the oscillatory nature of the solutions, we introduce radial and angular
coordinates $r$ and $\psi$
$$
\eqalign{\phi &= r \sin \psi\cr
y &= - 3^{- 1/4}\, r (1 - \cos \psi)}\, .\eqno(3.13)
$$
The resulting equations for $r$ and $\psi$ are
$$
\eqalign{\psi^\prime \, &= \omega\cr
r^\prime \, &= 0\, .}\eqno(3.14)
$$
where the frequency is $\omega = 2 \cdot 3^{1/4}$. The corresponding solutions
oscillate with frequency given by $H_0$ (which we expect to be Planck scale)
about Minkowski space.

\par
Based on the preceeding discussion of asymptotic solutions we see that there is
a separatrix$^{22)}$ in phase space dividing solutions which tend to $\phi =
\infty$ from those which oscillate or tend to $\phi =\, - \infty$. We observe
that for large $\vert H\vert$, the separatrix will asymptotically (and from the
right hand side on Fig. 2) approach the line of turning points given by $d \phi
/ dH = 0$.  From (3.1), it follows that for large $\vert H\vert$ the turning
points lie at
$$
\phi \simeq {1\over {\sqrt{3}}}\, .\eqno(3.15)
$$
For small values of $\phi$ and $H$, the separatrix is well to the right of the
line of turning points given by
$$
\phi \simeq 3^{1/4}\, {\vert H\vert\over {H_0}}\, .\eqno(3.16)
$$
\par
The above analysis of the phase space trajectories is an indication that in our
theory, Minkowski space is stable towards homogeneous perturbations.  As long
as the initial values of $\vert {\dot H}  \vert\, ,\, \vert {\dot \phi} \vert$
and ${\dot \phi} / \vert {\dot H} \vert$ are small, a solution starting close
to Minkowski space will remain close for all times.  The issue of stability of
Minkowski space towards inhomogeneous perturbations is an important unsolved
problem.
\par
We stress again that all the general features of the phase space analysis are
true for any potential $V (\phi)$ which satisfies the required asymptotic
conditions (2.19) and (2.21).  However, the results depend crucially on the
choice of the invariant $I_2$.
\par
Next, we include hydrodynamical matter with energy density
$$
\rho_m (t) = c\, a(t)^{-n}\, ,\eqno(3.17)
$$
where $n = 3$ for dust and $n=4$ for radiation. For the moment, we keep to a
collapsing spatially flat model.  In this case, equation (3.1) is unchanged
while equation (3.2) becomes
$$
{\dot \phi} = - 3H\phi + \sqrt{3} H - {1\over {2 \sqrt{3} H}}\, V - {8 \pi
G\over {\sqrt{3} H}}\, c\, a (t)^{-n}\, .\eqno(3.18)
$$
\par
With matter, phase space is three dimensional -- the third dimension being
$a(t)$.  In Fig. 3, we show the projection of some of the trajectories onto the
$(\phi (t)\, ,\, H(t))$ plane, for potential $V(\phi)$ given by (3.5).  All
trajectories have $8 \pi G\, c = 1$ and $a (t_0) = 10$, $t_0$ being the initial
time.  The main impression is that the trajectories look very similar to those
without matter in the asymptotic region.  We shall now explain why this is the
case.
\par
First, we note that as $\vert \phi \vert \rightarrow \infty$, the solutions
approach de Sitter space since ${\dot H} \rightarrow 0$.  Hence,
$$
a(t) \simeq e^{- \vert H\vert (t - t_0 )} \, a(t_0)\, .\eqno(3.19)
$$
Next, we combine (3.1) and (3.18) to obtain
for $\vert \phi \vert >> 1$
$$
{dH\over {d\phi}} \simeq {V^\prime\over {2 \sqrt{3}}}\> \left ( 3 H \phi +
{8\pi G c\over {\sqrt{3} H}}\> a (t)^{-n}\, \right )^{-1}\, .\eqno(3.20)
$$
Our model incorporates a very important feature: in the asymptotic de Sitter
region matter does not have an important effect on the geometry. The effective
gravitational constant which describes the influence of matter on the geometry
goes to zero as space-time approaches de Sitter space. In this sense the model
is asymptotically free.
\par
Some understanding of asymptotic freedom can be obtained by solving the $\phi$
and $H$ equations of motion (3.1) and (3.18) in the asymptotic region $\vert
\phi \vert >> 1$.  Equation (3.18) becomes
$$
\dot \phi \simeq 3 \vert H\vert \phi + {c\over {\vert H\vert }} a(t_0)^{-n}\>
e^{n \vert H\vert (t - t_0)}\eqno(3.21)
$$
(where we have incorporated the factor $8 \pi G / \sqrt{3}$ into the definition
of $c$).  From (3.21) it follows that $\phi (t)$ is a linear combination of the
homogeneous solution (3.4) and (assuming that $H \simeq {\rm const}$ ) the
inhomogeneous contribition $\phi_I (t)$
$$
\phi_I (t) = {c\over {n \vert H\vert^2}}\> a(t_0)^{-n}\>\left (e^{n \vert
H\vert (t - t_0)}\> -1\right )\eqno(3.22)
$$
For dust $(n = 3)$, both the homogeneous and inhomogeneous terms grow at the
same rate, and the coefficient of the inhomogeneous term is smaller.  Hence,
matter does not effect even the time dependence of the phase space
trajectories.     For radiation $(n = 4)$, $\phi_I (t)$ grows faster than
(3.4).  At sufficiently late times, therefore, it will dominate.  In this
period, however, we can (for potential (3.5)) solve the $H$ equation (3.6)  to
obtain
$$
H(t) \simeq H(t_1) - {2H_0^2\over {3n \vert H\vert}}\> \left ({\sqrt{3} n \vert
H\vert^2\over {c}}\right )^3\> a(t_0)^{3n}\> e^{-3n \vert H\vert (t_1 - t_0)}\,
,
\eqno(3.23)
$$
(where $t_1$ is some time $\gg t_0$ well into the asymptotic region)
which shows that the presence of matter does not effect the final value of the
curvature when starting the evolution in the asymptotic region.
\par
For small $\vert \phi \vert$, the presence of matter does have a significant
effect on the phase space trajectories.  As $a (t_0)$ decreases (or,
equivalently, $c$ and thus the matter energy density increase), the distortions
of the trajectories increase, as can be seen by comparing Figs. 3 and 4.
Figure 4 corresponds to a matter energy density which is ten times larger.
\par
Finally, we consider the effects of spatial curvature.  In this case, equations
(3.1) and (3.2) generalize to (see (2.26) and (2.27))
$$
\dot H = - {1\over {2 \sqrt{3}}} V^\prime + {k \over {a^2}}\eqno(3.24)
$$
$$
\dot \phi = - 3H \phi + {k\over {H a^2}} \phi - {1\over {2 \sqrt{3} H}} V +
\sqrt{3} H + \sqrt{3} {k \over {H a^2}} - {c\over {H a^n}}\, ,\eqno(3.25)
$$
where the constant $c$ is as in Eq. (3.21).
\par
In the case of the potential (3.5) and for $c = 0$, some resulting phase space
trajectories projected onto the $(\phi / H)$ plane are shown in Figs. 5 and 6.
For the trajectories of Fig. 6, the initial value of $a(t)$ was chosen to be
ten times smaller than in Fig. 5.  Hence, the effects of curvature are more
pronounced.
\par
Consider a sample trajectory of Fig. 5.  It starts out  with large initial
value of $a$.    The trajectory tends towards $\vert \phi \vert >> 1$ and $\dot
H \rightarrow 0$, as in the case $k = 0$.  Since $a(t)$ is now decreasing
almost exponentially, the role of curvature increases.  At a critical value of
$\phi$, the value of $\dot H$ becomes $0$.  This will occur when
$$
{1\over {2\sqrt{3}}} \>V^\prime\, (\phi (t)) = {k\over {a^2 (t)}}\,
.\eqno(3.26)
$$
Hence, the smaller the initial value of $a(t)$, the earlier (3.25) will be
satisfied (compare Figs. 5 and 6).  At a similar time, the curvature terms also
start to dominate in Eq. (3.24).  Therefore, as is obvious from the $k$
dependent terms in (3.24), $\phi (t)$ will rapidly decrease, as will $\vert H
(t)\vert$.  At some finite and negative value of $\phi\, ,\, H(t)$ vanishes.
Thereafter, the Universe reexpands.  The evolution of this model for small
$a(t)$ resembles a de Sitter bounce.
\par
Note that all solutions are nonsingular.  In particular, the solutions can be
integrated through the point when $H=0$ (when terms on the right hand side of
(3.29) become infinite).
\par
In conclusion, we have constructed a higher derivative modification of
Einstein's theory in which all homogeneous and isotropic solutions are
nonsingular.  Without curvature (i.e. for $k=0$), the solutions either are
periodic about Minkowski space or else converge to a $k=0$ de Sitter solution.
For $k\neq 0$ the solutions which do not remain close to Minkowski space go
through a de Sitter bounce and are future extendible to $t = \infty$.  In
addition, we have shown that our model is asymptotically free in the sense that
the effective coupling of matter to gravity goes to zero as the curvature
increases.
\chapter{Nonsingular Universe with Limiting Curvature}
\par
Now we turn to the discussion of the full model in which the LCH is
implemented, the model given by the action (2.24), in which for a homogeneous
and isotropic metric the equations of motion reduce to (2.25 -- 2.27).  We
include hydrodynamical matter with energy density given by (3.17).
\par
In the general case $(k \neq 0 \>{\rm and}\>c\neq0)$, the phase space of the
model is three dimensional:  $\phi_1 (t)\, ,\, \phi_2 (t)$ and $a (t)$.  For
$k=0$ and $c=0$, the dependence on $a(t)$ drops out and the phase space  can be
reduced to the two dimensional $\phi_1 /\phi_2$ diagram.  The first order
equations of motion in phase space are found by combining equations (2.25 --
2.27).  To derive the equation for $\phi_1 (t)$, we differentiate (2.25) with
respect to $t$ and use (2.26) to substitute for $\dot H$ to obtain
$$
{\dot \phi}_1 = - 4 \sqrt{3}\> {HV_2^\prime\over {V_1^{\prime\prime}}}\,
.\eqno(4.1)
$$
The equation of motion for $\phi_2$ is (2.27):
$$
{\dot \phi}_2 = -3H\phi_2 + {k\over {Ha^2}} \phi_2 + {1\over {\sqrt{3} H}}\left
(3H^2 (1-2\phi_1\, ) + 3 {k\over {a^2}} (4\phi_1 - 1) - {1\over 2} (V_1 + V_2)
- {c\over {a^n}}\right )\eqno(4.2)
$$
where $H$ can be expressed in terms of $\phi_1$ and $a$ via (2.25).  From
(4.1), (4.2) and (2.25) it is obvious that for $k =c=0$ the $a(t)$ dependence
disappears.
\par
In the case $k=c=0$  we may use (2.25) to get
$$
{d\phi_2\over {d\phi_1}} = - {V_1^{\prime\prime}\over {4V_2^\prime}} \left [ -
\sqrt{3} \phi_2 + (1-2\phi_1) - {2\over {V_1^\prime}} (V_1 + V_2)\right ]\,
,\eqno(4.3)
$$
the key equation for the following phase space analysis.
\par
For all potentials $V_1(\phi_1)$ and $V_2(\phi_2)$ with asymptotical behavior
$$
V_i \propto \phi_i^2  \,\,\,\,\,\, \phi_i \ll 1 \eqno(4.4)
$$
and
$$
V_1 \propto \phi_1 - {\rm ln} \phi_1 + O({1 \over {\phi_1}}) \,\,\,\, \phi_1
\gg 1 \eqno(4.5)
$$
$$
V_2 \propto const + O({1 \over {\phi_2}}) \,\,\,\, \phi_2 \gg 1 \eqno(4.6)
$$
the phase diagrams have the same features as depicted schematically in Fig. 8
for spatially collapsing Universes without matter.
The numerical solutions depicted in Fig. 7 were obtained for the particular
choice of potentials (which satisfy the asymptotic conditions of (4.4 - 4.6))
$$
V_1 (\phi_1) = 12 H_0^2 {\phi_1^2\over {1+\phi_1}} \left (1 - {\ell n
(1+\phi_1)\over {1+\phi_1}}\right )\eqno(4.7)
$$
$$
V_2 (\phi_2) = 2 \sqrt{3} H_0^2 {\phi_2^2\over {1+\phi_2^2}}\, .\eqno(4.8)
$$
The presence of the logarithmic  term in (4.4) will be justified shortly.
\par
We can identify four classes of trajectories.  Note that by (2.26), $\vert
\phi_2 \vert \rightarrow \infty$ implies that the evolution approaches de
Sitter space. The first class of trajectories starts in the de Sitter phase at
$\phi_2 \rightarrow - \infty$ and evolves to de Sitter at $\phi_2 \rightarrow
\infty$.  For small initial values of $\phi_1$, trajectories starting at
$\phi_2 = - \infty$ reach a turning point and return to $\phi_2 = - \infty$.
The third class of trajectories are periodic solutions about Minkowski
space--time $(\phi_1 = \phi_2 = 0)$.  Finally, trajectories starting with small
$\phi_1$ and $\phi_1 / \phi_2$ with $\phi_2$ positive evolve towards de Sitter
solutions at $\phi_2 = \infty$.  There are two separatrices dividing phase
space into regions corresponding to the four above classes (see Fig. 8).
\par
Note that in order to prevent solutions starting with $\phi_1 >> 1$ and $\phi_2
\simeq 0$ from escaping to $\phi_1 = \infty$ at $\phi_2 < 1$ in finite time -
such solutions would violate the LCH and would lead to singularities in higher
order curvature invariants - it was necessary to add the logarithmic correction
term to $V_1 (\phi_1)$.
\par
Phase space is the half plane $\phi_1 \geq 0$.  Negative values of $\phi_1$ are
unphysical since by (2.25), and using the small $\phi_1$ asymptotic form of
$V_1 (\phi_1)$, they would correspond to imaginary values for $H(t)$.  This
half plane can be divided into four regions:  in Region A, $\phi_1 >> 1$ and
$\vert \phi_2 \vert >> 1$, in Region B, $\phi_1 >> 1$ and $\vert \phi_2 \vert
<< 1$, in Region C, $\phi_1 << 1$ and $\vert \phi_2 \vert << 1$ and in Region
D, $\phi_1 << 1$ and $\vert \phi_2 \vert >> 1$.  We will analyze the phase
space trajectories in each of the above regions, focusing on three features:
the asymptotic expressions for $d\phi_2 / d \phi_1$ (which give the tangent
vectors to the trajectories), the separatrices, and the equations for the
trajectories.  To concretize the discussion, we use the potentials (4.7) and
(4.8).  However, except in Region B, the asymptotical solutions are independent
of the specific choice of potentials.
\par
In Region A, Equation (4.3) becomes
$$
{d\phi_2\over {d\phi_1}} \simeq {\sqrt{3}\over {4}}\> {\phi_2^4\over
{\phi_1^2}} \left [ \sqrt{3} + 4 {\phi_1\over {\phi_2}}\right ]\, .\eqno(4.9)
$$
The direction of the tangent vectors is sketched in Fig. 8.  Arrows indicate
the direction of increasing time and are obtained by inspecting (4.1) and (4.2)
directly.  By inspecting the tangent vectors it is clear that all solutions in
the upper region $\phi_2 > 1$ quickly approach de Sitter space ($\vert \phi_2
\vert \rightarrow \infty$ implies de Sitter space).  In the lower region
$\phi_2 < - 1$ there are two domains separated by a separatrix which for
$\phi_1 \gg 1$ and $\vert \phi_2 \vert \gg 1$ is close to the line of turning
points where $d\phi_2 /d\phi_1 = 0$, its equation being given by
$$
{\phi_2\over {\phi_1}} = - {4\over {\sqrt{3}}}\eqno(4.10)
$$
(see Fig. 8).  To the right of the separatrix, trajectories correspond to
solutions starting out in a de Sitter phase.  To the left of the line given by
(4.10), we have $d \phi_2 /d\phi_1 > 0$ and trajectories go off to de Sitter
space at $\phi_2 \rightarrow - \infty$.  In all cases, de Sitter space is
reached at finite $\phi_1$ values.  This is seen by explicitly integrating
(4.9).  In the region where the first term on the right hand side of (4.9)
dominates we have
$$
\phi_1 \simeq c - {2\over {9}}\> \phi_2^{-3}\, ,\eqno(4.11)
$$
while in the domain where the second term dominates the approximate solution is
$$
\phi_1 \simeq {1\over {{9\over 4}\> \phi_2^3 +c}}\, \eqno(4.12)
$$
($c$ is a constant of integration).
Note that all of the solutions starting in Region A start in de Sitter space
and end up in de Sitter space.
\par
In Region B, the tangents in phase space are given by
$$
{d\phi_2\over {d\phi_1}} \simeq \sqrt{3}\> {1\over {\phi_1 \phi_2}}\eqno(4.13)
$$
 which integrates to
$$
\phi_1 = c \exp \left\{ {1\over {2\sqrt{3}}}\> \phi_2^2\right\}\, .\eqno(4.14)
$$
The tangent vectors are again sketched in Fig. 8.  From (4.14) it follows that
trajectories leave Region B at a finite value of $\phi_1$.  They enter Region A
and hence asymptotically approach de Sitter space.
\par
In Region C, Equation (4.3) becomes
$$
{d\phi_2\over {d\phi_1}} \simeq - {\sqrt{3}\over {2 \phi_2}} \left [ 1 -
3\phi_1 - {1\over {\sqrt{12}}}\> {\phi_2^2\over {\phi_1}} \right ]\,
.\eqno(4.15)
$$
The separatrix in the upper half planes is close to the line of turning points
$d\phi_2 / d\phi_1 = 0$ for large values of $\phi_i$:
$$
\phi_2 \simeq\,  \pm\, 12^{1/4} \phi_1^{1/2} (1 - 3 \phi_1 )^{1/2}\,
.\eqno(4.16)
$$
Where the first term dominates, the trajectories obey
$$
\phi_1 \simeq - {1\over {\sqrt{3}}}\> \phi_2^2 + c\, .\eqno(4.17)
$$
{}From the sketch of Fig. 8 it is clear that the trajectories which pass
through $\phi_1 = \phi_2 = 0$ with ${\dot \phi}_2 /{\dot \phi}_1 (\phi_1 =
\phi_2 = 0)$ not too large correspond to periodic motion about Minkowski space.
 This -- as in the model of Section 3 -- is an indication that Minkowski space
is stable in our theory towards homogeneous perturbations.
\par
Finally, in Region D the equation for the tangent vector is
$$
{d\phi_2\over {d \phi_1}} \simeq + {\sqrt{3}\over 2} \phi_2^3 \left [ \sqrt{3}
\phi_2 + {1\over {2 \sqrt{3} \phi_1}}\right ]\, .\eqno(4.18)
$$
There is a separatrix which is (for large $\phi_i$) approximately described by
$$
\phi_2 \simeq - {1\over {6 \phi_1}}\, .\eqno(4.19)
$$
To the right of this line, the trajectories are given by
$$
\phi_1 \simeq c - {2\over {9}}\> \phi_2^{-3}\, ,\eqno(4.20)
$$
to the left by
$$
\phi_1 \simeq c\, e^{- 2 \phi_2^{-2}}\, .\eqno(4.21)
$$
\par
In conclusion, all solutions are either periodic about Minkowski space or are
asymptotically de Sitter.  All solutions can be extended to $t = \pm\, \infty$,
and hence there are no singularities.
\par
As in the model of Section 3, including matter does not effect the asymptotic
solutions.  The coupling between matter and gravity is asymptotically free also
in the theory with action (2.24).  However, including matter changes the nature
of solutions starting near Minkowski space.  These solutions now approach de
Sitter space (see Fig. 9).  This result is not surprising, since also in
Einstein gravity, Minkowski space is not a solution of the field equations in
the presence of matter.
\par
The projection of some phase space trajectories onto the $(\phi_1 \, ,\,
\phi_2)$ plane in a model with $k \neq 0$ but $c=0$ are shown in Fig. 10.  Like
in the single field model of Section 3, the trajectories initially evolve as
for $k=0$ towards de Sitter space.  Hence, for finite $\phi_1\, ,\, \phi_2$
becomes very large.  Eventually, however, the curvature terms become important,
$\phi_2$ reaches a turning point and rapidly (within time period $H_0^{-1}$)
relaxes to zero (for finite value of $\phi_1$).  As is obvious from Fig. 11,
the rapid decrease in $\phi_2$ corresponds to the de Sitter bounce during which
$H$ changes sign.
\chapter{Conclusions and Discussion}

\par
We have constructed a theory of gravity in which all homogeneous and isotropic
solutions (not only special solutions as in some other models$^{23)}$) are
nonsingular, regardless of the matter content of the Universe.  Our effective
action for gravity contains higher derivative terms which modify the Einstein
action at high curvatures.  Such terms are expected to be important near the
Planck curvature in any fundamental theory such as quantum gravity or string
theory.
\par
Most higher derivative gravity theories have much worse singularity properties
than Einstein gravity.  We use a particular construction based on implementing
the  ``Limiting Curvature Hypothesis'' to obtain a class of models without
singularities.  We discussed two models, one in which all curvature invariants
are bounded and all solutions except those periodic about Minkowski space
asymptotically approach de Sitter space (Section 4), and a simpler model
without limiting curvature (Section 3).
\par
The theory presented in this paper is ``asymptotically free" in the sense that
the coupling of matter to gravity goes to zero as the curvature approaches its
limiting value (similar features have been discussed by Linde$^{24)}$ under the
name ``gravitational confinement").
\par
When applied to an expanding Universe, our theory implies that it started out
in a de Sitter phase with scale factor $a(t) = e^{Ht}$ (for $k=0$) or else (for
$k=1$) it emerged from a de Sitter bounce.  In particular, there was a period
of inflation driven by gravity.  This is no surprise as it is well
known$^{25)}$ that higher derivative gravity theories often produce inflation.
\par
Note that the property of asymptotic freedom might also justify using the
effective action approach to gravity until Planck curvatures. Asymptotic
freedom will also play an important role in controlling nonlocal terms. For
example, nonlocal terms due to particle production may be expected to vanish in
the asymptotic regions of phase space.
\par
Our action is constructed by adding two Lagrange multiplier terms (and their
corresponding potentials) to the Einstein action.  Each Lagrange multiplier is
coupled to a curvature invariant.  The role of the first Lagrange multiplier is
to limit the curvature, the role of the second one $(\phi_2)$ is to force
space--time to be de Sitter at large curvature.  For a homogeneous and
isotropic model, it was sufficient to couple $\phi_2$ to the invariant $I_2 = 4
R_{\mu \nu} R^{\mu \nu} - R^2$, since in this case $I_2 = 0$ singles out de
Sitter space.
\par
However, for an anisotropic cosmology, we must extend the invariant $I_2$ by
including a term which effects the anisotropy.  In a subsequent paper$^{13)}$
(see also Ref. 26) we show that $I_2 = + R_{\mu \nu} R^{\mu \nu} - R^2 + C^2$
is an appropriate
invariant.  This invariant also works for a spherically symmetric metric.
Thus, in a model like the one presented here, but with the new $I_2$, we are
able to show that there will be no singularities inside the black hole
horizon$^{13)}$.
\par
Open questions include the generalization of our model to general inhomogeneous
metrics, and a full stability analysis.
\medskip
\noindent \undertext{Acknowledgement}
\par
For useful discussions we are grateful to B. Altshuler, A. Linde, M. Markov, M.
Mohazzab and M. Trodden.  This work has been supported in part by the U.S.
Department of Energy under grant No. DE-AC02-76-ER03130 (Task A), by an Alfred
P. Sloan Foundation fellowship to R.B., and by the Swiss National Science
Foundation (V.M.).

\bigskip
\REF\one{R. Penrose, {\it Phys. Rev. Lett.} {\bf 14}, 57 (1965);\hfill\break
S. Hawking, {\it Proc. R. Soc. London}, {\bf A300}, 182 (1967).}
\REF\two{C. Misner, K. Thorne and J. Wheeler, Gravitation (Freeman, New York,
1972).}
\REF\three{R. Brandenberger and C. Vafa, {\it Nucl. Phys.} {\bf B316}, 391
(1989);\hfill\break
A. Tseytlin and C. Vafa, {\it Nucl. Phys.} {\bf B372}, 443 (1992);\hfill\break
G. Veneziano, {\it Phys. Lett.} {\bf B265}, 287 (1991);\hfill\break
M. Gasperini and G. Veneziano, {\it Phys. Lett.} {\bf B277}, 256
(1992);\hfill\break
M. Gasperini and G. Veneziano, { ``Pre--Big-Bang in String Cosmology'', CERN
preprint TH6572/92 (1992).}}
\REF\four{C. Clarke, {\it Comm. Math. Phys.} {\bf 41}, 65 (1975).}
\REF\five{J. Narlikar and T. Padmanabhan, {\it Ann. Phys. (N.Y.)} {\bf 150},
289 (1983).}
\REF\six{M. Markov, {\it Pis'ma Zh. Eksp. Teor. Fiz.} {\bf 36}, 214
(1982);\hfill\break
M. Markov, {\it Pis'ma Zh. Eksp. Teor. Fiz} {\bf 46}, 431 (1987);\hfill\break
V. Ginsburg, V. Mukhanov and V. Frolov, {\it Zh. Eksp. Teor. Fiz.} {\bf 94}, 1
(1988).}
\REF\seven{R. Penrose, in ``General Relativity:  An Einstein Centenary'', ed.
by S. Hawking and W. Israel (Cambridge Univ. Press, Cambridge, 1979).}
\REF\eight{V. Frolov, M. Markov and V. Mukhanov, {\it Phys. Lett.} {\bf B216},
272 (1989);\hfill\break
V. Frolov, M. Markov and V. Mukhanov, {\it Phys. Rev.} {\bf D41}, 383
(1990);\hfill\break
D. Morgan, {\it Phys. Rev.} {\bf D43}, 3144 (1991).}
\REF\nine{S. Hawking, {\it Phys. Rev.} {\bf D14}, 2460 (1976).}
\REF\ten{C. Callen, S. Giddings, J. Harvey and A. Strominger, {\it Phys. Rev.}
{\bf D45}, R1005 (1992).}
\REF\eleven{V. Mukhanov and R. Brandenberger, {\it Phys. Rev. Lett.} {\bf 68},
1969 (1992).}
\REF\twelve{B. Altshuler, {\it Class. Quant. Grav.} {\bf 7}, 189 (1990).}
\REF\thirteen{R. Brandenberger, M. Mohazzab, V. Mukhanov, A. Sornborger and M.
Trodden, in preparation (1993).}
\REF\fourteen{A. Guth, {\it Phys. Rev.} {\bf D23}, 347 (1981).}
\REF\fifteen{G. 't Hooft and M. Veltman, {\it Ann. Inst. H. Poincar\'e} {\bf
20}, 69 (1974);\hfill\break
A. Barvinsky and G. Vilkovisky, {\it Phys. Rep.} {\bf 119}, 1 (1985).}
\REF\sixteen{N. Birrell and P. Davies, ``Quantum Fields in Curved Space'',
(Cambridge Univ. Press, Cambridge, 1992).}
\REF\seventeen{J. Scherk and J. Schwarz, {\it Nucl. Phys.} {\bf B81}, 118
(1974);\hfill\break
T. Yoneya, {\it Prog. Theor. Phys.} {\bf 51} 1907 (1974);\hfill\break
C. Lovelace, {\it Phys. Lett.} {\bf 135B}, 75 (1984);\hfill\break
E. Fradkin and A. Tseytlin, {\it Nucl. Phys.} {\bf B261}, 1 (1985).}
\REF\eighteen{M. Markov and V. Mukhanov, {\it Nuovo Cimento} {\bf 86B}, 97
(1985).}
\REF\nineteen{B. Whitt, {\it Phys. Lett.} {\bf B145} 176 (1984).}
\REF\twenty{M. Born and L. Infeld, {\it Proc. R. Soc. London} {\bf A144}, 425
(1934).}
\REF\twentyone{R. Brandenberger, R. Kahn and W. Press, {\it {Phys. Rev.}} {\bf
D28}, 1809 (1983).}
\REF\twentytwo{see e.g. V. Arnold, {\it Ordinary Differential Equations} (MIT
Press, Cambridge, 1973).}
\REF\twentythree{T. Ruzmaikina and A. Ruzmaikin, {\it JETP Lett.} {\bf 30},
372 (1970);\hfill\break
H. Nariai and K. Tomita, {\it Prog. Theor. Phys.} {\bf 46}, 776
(1971);\hfill\break
M. Giesswein and E. Steernwitz, {\it Acta Phys. Austriaca} {\bf 41}, 41
(1975);\hfill\break
V. Gurovich, {\it Zh. Eksp. Teor. Fiz.} {\bf 73},369 (1977);\hfill\break
J. Barrow and A. Ottewill, {\it J. Phys.} {\bf A16}, 2757 (1983).}
\REF\twentyfour{A. Linde, {\it Rep. Prog. Phys.} {\bf 47}, 925 (1984).}
\REF\twentyfive{A. Starobinski, {\it Phys. Lett.} {\bf B91}, 99 (1980).}
\REF\twentysix{B. Altshuler, in ``Proc. of lst Int. A.D. Sakharov Conference'',
Moscow, May 1991 (NOVA Science Publishing, Inc., New York, 1992).}
\refout
\endpage
\centerline{\bf Figure Captions}

\item{Fig. 1:} Penrose diagram of an eternal black hole in Einstein gravity (a)
and in the nonsingular Universe theory (b). The singularities (S) are replaced
by de Sitter phases (dS) which couple to Friedmann Universes (FRW). The
horizons (H) are not affected.

\item{Fig. 2:} Phase-space diagram $(\phi, H)$ - arrows indicating the
direction of time evolution - for a spatially flat Universe without limiting
curvature and with no matter $(k = c = 0)$. Generated using the potential
(3.5).

\item{Fig. 3:} A projection onto the $(\phi, H)$ plane of the three dimensional
phase-space diagram $(\phi, H, a)$ - arrows again indicating direction of time
evolution - for a spatially flat Universe without limiting curvature but with
matter $(k = 0, c \neq 0)$. Generated using the potential (3.5) with initial
condition $a(t_0) = 10$.

\item{Fig. 4:} Phase-space diagram as in Fig. 3, but with $a(t_0) = 1$.
Therefore, the initial matter energy density is larger than for the
trajectories of Fig. 3.

\item{Fig. 5:} A projection onto the $(\phi, H)$ plane of the three dimensional
phase-space diagram $(\phi, H, a)$ in a closed $(k = 1)$ Universe without
limiting curvature and in the absence of matter $(c = 0)$. Potential (3.5) was
used, and $a(t_0) = 10$ was chosen as initial condition.

\item{Fig. 6:} Same as in Fig. 5, but with initial condition $a(t_0) = 1$.
Notice the different scales on the axes.

\item{Fig. 7:} Phase-space diagram for the spatially flat $(k = 0)$ Universe
with limiting curvature based on the potentials (4.7) and (4.8). There is no
matter $(c = 0)$.

\item{Fig. 8:} Sketch of the generic phase-space diagram for a two field model
with $k = c = 0$ and potentials satisfying the asymptotic conditions (4.4) -
(4.6). Lines with arrows indicate phase space trajectories, arrows pointing in
the direction of increasing time. Separatrices are shown as dashed lines. With
$A, B, C$ and $D$ we denote the asymptotic regions of phase space discussed in
the text.

\item{Fig. 9:} A projection onto the $(\phi_1, \phi_2)$ plane of the three
dimensional phase-space diagram $(\phi_1, \phi_2, a)$, for a two field model
which is spatially flat but contains matter. The potentials used are (4.7) and
(4.8).

\item{Fig. 10:} The same for a two field model without matter but including
spatial curvature $(k \neq 0)$.

\item{Fig. 11:} Trajectories in the $(I_2, H)$ plane for the same model as in
Fig. 10.
\end